# The Use of Learning Management Systems for Self-paced Learning: The Case at a South African Public Access Centre

Guidance Mthwazi[1[0000-0002-4078-1214]] Meke Kapepo[1[0000-0001-7829-4852]] and Jean-Paul Van Belle[1[0000-0002-9140-0143]]

[1]University of Cape Town, Rondebosch, South Africa
meke.kapepo@uct.ac.za

**Abstract.** This study investigates the use of a Learning Management System (LMS) to support self-paced learning at a South African Public Access Centre (PAC), using the I-CAN Centre as a case study. Through semi-structured interviews with thirty-eight learners and thematic analysis, the research explores opportunities and challenges associated with LMS adoption. Findings reveal that PACs play a critical role in promoting ICT skills and digital inclusion, offering learners flexible access to learning resources and fostering empowerment. While LMS use enhances convenience and supports blended learning as the preferred approach, persistent challenges, such as poor connectivity, outdated infrastructure, and unclear course instructions, limit its effectiveness. These findings highlight the need for infrastructural upgrades and user-centric design to optimise LMS implementation in community-based learning environments.

**Keywords:** Learning Management System (LMS), Public Access Centres (PACs), I-CAN Centre, Self-paced learning, Blended learning, Learners.

## 1 INTRODUCTION

The access, adoption, and use of Information and Communication Technologies (ICTs) have significantly impacted the education sector worldwide. Through their utilisation, education institutions have improved teaching and learning standards and helped develop practical pedagogies to support learning processes [1]. Access to ICTs is essential for the innovation and development of an economy; however, in countries with developing economies, access remains unequal despite the exponential growth of ICT networks, services, and applications [2]. But South Africa has a relatively very low internet penetration rate, an issue more pronounced within underserved communities [3].

The South African government has established various programmes to rapidly increase internet accessibility and eradicate the digital divide, especially within underserved communities [2]. Part of such initiatives is the establishment of facilities, are Public Access Centres (PACs), which offer ICT services such as e-learning, e-government services, and other social and economic development services [4]. Learning Management Systems (LMS) form part of these interventions. LMS augments conventional face-to-face teaching and learning by facilitating learner participation and improving information sharing and communication between learners and instructors without location and time restrictions [5].

Despite this, there is a knowledge gap in the literature regarding LMS use at PACs. Studies have examined the adoption, challenges, and acceptance of LMS use in higher education [5], [6]. Similarly, [7] explored the challenges that impact e-learning systems, but within a Covid-19 environment. As such, to augment the ongoing discussion, this study, examined the use of an LMS at a Public Access Centre, a development that can enhance its use within a context that is easily accessible to the public. Additionally, the study explored the opportunities and challenges experienced by a diversity of learners (given that it is in a public access scenario) when using LMS for self-paced learning. To execute that, the following were the objectives of the study: to describe the use of the LMS to facilitate self-paced learning courses at the I-CAN Centre; and to explain the challenges and opportunities experienced by self-paced learners when using the LMS at the I-CAN Centre.

This paper aims to demonstrate how Public Access Centres (PACs) can leverage Learning Management Systems (LMS) to provide affordable, accessible, and flexible learning opportunities that develops and enhance ICT skills and empower underserved communities. It also highlights the infrastructural and usability challenges that must be addressed for improved implementation. The remainder of the paper is categorised into four more sections namely, a literature review, methodology, findings and discussion section, and finally, concluding remarks.

## 2 LITERATURE REVIEW

### 2.1 LMS use, opportunities and challenges

The use of LMS as an innovative medium to deliver education has challenged the traditional teaching approach, allowing learners to obtain formal education in a non-conventional learning environment such as PACs. Access to a LMS at PACs can enrich educational experiences, encouraging learners to become active participants in developing skills, competencies and capability to solve problems [6]. LMS offers features that cater to the educational needs of a 21st-century learner and instructor; these include accessing learning content using any device and connecting anywhere and at any time using wireless networks. The mElimu system, for example, a LMS software for e-learning that works across multiple devices including tablets and laptops [8], can facilitate learning activities and manages related learning content in an interactive learning environment. Its revolutionary digital learning platform allows students to learn anywhere, anytime, at their convenience [8], and provides various functionalities, including the ability for learners to view course completions, course statuses, course certification, and manage course content.

LMS applications have, over the years, become advanced, providing many features that complement education needs and opportunities. These features include access without the restriction of location and time, enabling learners to access learning content without being at a physical location [9]. For example, PACs with many learners enrolled in courses at their facilities would provide learners access to their LMS content regardless of where they are. Accessibility to the LMS is another advantage; the advancement of technology has improved how learners access the application. The mElimu system has been at the forefront of mobile education, enabling users to access e-learning content using any device. Learners and instructors can access numerous resources to facilitate collaboration and create innovative learning environments [10].

However, despite these opportunities, technology determinism has often taken a top-down approach for decades. This, because of the digital first approach, has led to implementation failure and a lack of user acceptance [11]. Some constraints are a lack of good governance, lack of user centricity, technical infrastructure, ICT knowledge and finances [7]. Other limitations include readiness to use an LMS. The lack of preparedness occurs when a learner lacks the confidence to perform a specific task successfully due to limited computer skills and technical support [5]. Furthermore, having an incompetent course instructor could impede the adoption and use of an LMS [11]. Online assessments also pose various challenges due to a lack of expertise in conducting checks effectively and guarding against plagiarism, a development that saw an exponential rise during the COVID-19 Pandemic phase [12]. Another critical factor is that lack of managerial support can restrict the promotion and use of LMS in PACs. Organisational support is integral in influencing employee attitude, allocating resources and managing operational efficiencies in LMS use [5].

### 2.2 Public Access Centres [PACs]

The South African government in the late 1990s established various programmes to stimulate economic and social inequalities in the country [13]. They included establishing facilities such as PACs or telecentres, libraries, and information kiosks. These facilities were launched with an explicit mandate to provide public access to information and developmental e-services to underserved communities, including access to free or reduced-fee government services and other ICT4D programmes. With the high youth unemployment in South Africa, implementing these programmes may enable them to develop technical and employable skills [3]. While access to these PACs may not be entirely free of charge, accessibility fees are relatively low to accommodate the have-nots in the communities [2].

However, PACs in South Africa and other developing economies have not made a significant impact in transforming communities socially and economically [2]. The challenge PACs face is their inability to align programmes within the context of the communities it intends to serve. Further limitations include inadequate adoption of the technologies, often linked to a learner’s educational level or experience using these systems. Another limitation is the lack of affordable access to the internet, decreasing the usage of PACs’ ICT programmes. These limitations contribute to digital inequality, differentiating those with the skills and resources to use the e-learning technologies optimally from those without.

### 2.3 Using PACs to enhance approaches to learning

The effective adoption of learning approaches requires PACs to implement strategies that foster engagement and support diverse learning contexts. These strategies include facilitating conducive learning environments, managing learner interactions, and overseeing learning processes [14]. Implementing a learning approach also demands that learners actively monitor, regulate, and adapt their behaviours and actions in response to specific learning situations. A well-integrated combination of face-to-face (F2F) and online learning has been shown to enhance the quality of education and promote interactive learning outcomes. Beyond the traditional F2F model, two additional modes of learning have emerged: blended learning and self-paced learning. This study focuses on one of these alternative approaches, examining its role in improving learner engagement and overall educational effectiveness.

Blended learning approaches provide an education environment that combines e-learning technologies and F2F learning to attain learning instruction online and in a facilitated learning setting [14]. This learning approach allows learners to take responsibility for their learning processes and utilise external sources to support their learning activities. At the same time, it transforms an instructor's role from teaching to facilitator, allowing learners to explore the learning content independently and encouraging independent problem-solving and a convenient way of studying [15]. The focus of this study is self-paced learning, and this refers to the learners' perception of their independent education, sense of responsibility for their knowledge, and initiative to learn. The learning approach requires learners to engage with learning processes, acquire information, plan, evaluate learning activities, and improve learning processes, participation, and performance [14]. Self-paced learning allows learners to receive support and feedback based on their required needs, work through the feedback at their own pace, and request additional assistance when the need arises [16].

While literature has indicated that PACs can facilitate more of self-paced learning than blended, this study embarked on describing the use of the LMS to facilitate self-paced learning at the I-CAN Centre (PAC). It further examined the challenges and opportunities experienced by self-paced learners when using the LMS at the I-CAN Centre. The use of a LMS at a PAC was found to be a knowledge gap that has not been examined in depth in the extant literature.

## 3 RESEARCH METHODOLOGY

A research philosophy that a researcher adopts reveals their positionality and assumptions on the development of knowledge in a field [17]. Assumptions shape the understanding of the research question, method, and interpretation of the research findings [18]. We adopted the interpretivism research philosophy for the research study. Interpretivism recognises that interpreting research material, data, and values plays a significant role in research [19]. The study considered the different views of participants and how they perceived the world around them. Geographical, gender, ethnic, experiential, and social differences among the learners resulted in different perspectives on using LMS at the I-CAN Centre by the diverse group of learners interviewed.

An inductive approach to theory was adopted for the research study. Given the explanatory nature of the research questions which sought first to describe the use of LMS to facilitate self-paced learning and then examine how learners experience challenges and opportunities when using the LMS to facilitate self-paced learning at the I-CAN Centre, a descriptive purpose was initially employed, aligning with the first research objective. This was preceded with an explanatory purpose which aligns to the second objective [20]. Semi-structured interviews were conducted with a sample of learners to gather evidence of their experience using LMS at the I-CAN centre. The interviews were versatile and flexible enabling the researchers to improvise follow-up questions depending on the responses received from the participants [21]. Initial questions (e.g., about frequency of use of LMS per week) manifested descriptive responses, while follow up questions (e.g., how they found the experiences to be) aided the explanatory part of the study.

A case study approach was adopted. A case study approach allows for a detailed investigation that provides an analysis to unique or underexplored contexts and processes involved in the phenomenon under study [17]. The observation entailed in a case study enables a researcher to study and examine different aspects, view the processes within an environment, and use a researcher's in-depth capacity to understand some phenomenal behaviour. Learners who had classes during the scheduled interview week formed part of the sample. Prior to interviews, ethical clearance was sought from the university, and ethics clearance was also granted by the director of the I-CAN Centre before the research commenced. Participants consented to participation and were voluntarily given an option to exit the research at any point during the interview. Learners were approached for an interview if they were enrolled in a course at the I-CAN Centre (our case study) and had used the LMS.

Of the 38 purposively sampled learners who met the criteria, 14 were female, and 24 were male. The learners were between 18 and 35 years old, with the majority being between 18 and 24. Most of the learners had completed matric when enrolled in a course. The learners could select from the varied courses offered at the I-CAN Centre. Seven (7) of the learners enrolled in the Graphical Design (C1) course, 12 in the IT Technical Engineering (C2) course, 4 in the IT Technician (C3) course, 8 in the IT Web Engineering (C4) course, and 7 in the Office Administration (C5) course. Each learner provided a different perspective and experience using the I-CAN Centre's LMS. The interview guide was semi-structured developed based on the findings from the literature review, but with additional questions that evolved during the research process. Interviews were recorded and transcribed (Otter.ai) and the analysis was done using Nvivo.

The researchers used thematic analysis to analyse the data collected. Thematic analysis is used to identify and organise findings into patterns of themes across the data. It furthermore assists the researchers in extracting relevant information to determine relationships between the constructs and to create a logical data flow to understand the study better. The thematic analysis was guided by Braun and Clarke's [22] widely used guide for thematic analysis.

# 4 PRESENTATION AND DISCUSSION OF FINDINGS

This section is dedicated to representation and presentation of the study's findings. It further incorporated the discussion sections alongside the findings to describe and explain the findings in detail. The findings are categorised into descriptive findings which mostly presented the statistical implications of the study, as well as the narrative findings, which details the explanatory findings. Finally, the last section consolidates both the descriptive and explanatory findings.

## 4.1 Descriptive findings on LMS use

Analysis of the first research question on LMS use for self-paced learning revealed key descriptive insights. All participants reported prior experience with an LMS or similar systems, which was a criterion for sampling. This information helped assess learners' familiarity with digital platforms before enrolling at the I-CAN Centre. Further exploration examined learners' ability to access the LMS at home and manage learning activities beyond the Centre. Because the Centre primarily uses the mElimu LMS, most of the discussion in what follows refers to the mElimu system.

Likewise, access to computers and the internet at home was a follow-up theme analysed to understand whether the learners had access to the technical infrastructure to continue accessing the learning activities on the LMS while at home. One of the attractive features of LMS was its accessibility at any time and from anywhere if users have the technical infrastructure to connect to it. This finding helped us understand whether the learners had the necessary infrastructure to take advantage of, which could stimulate self-paced learning.

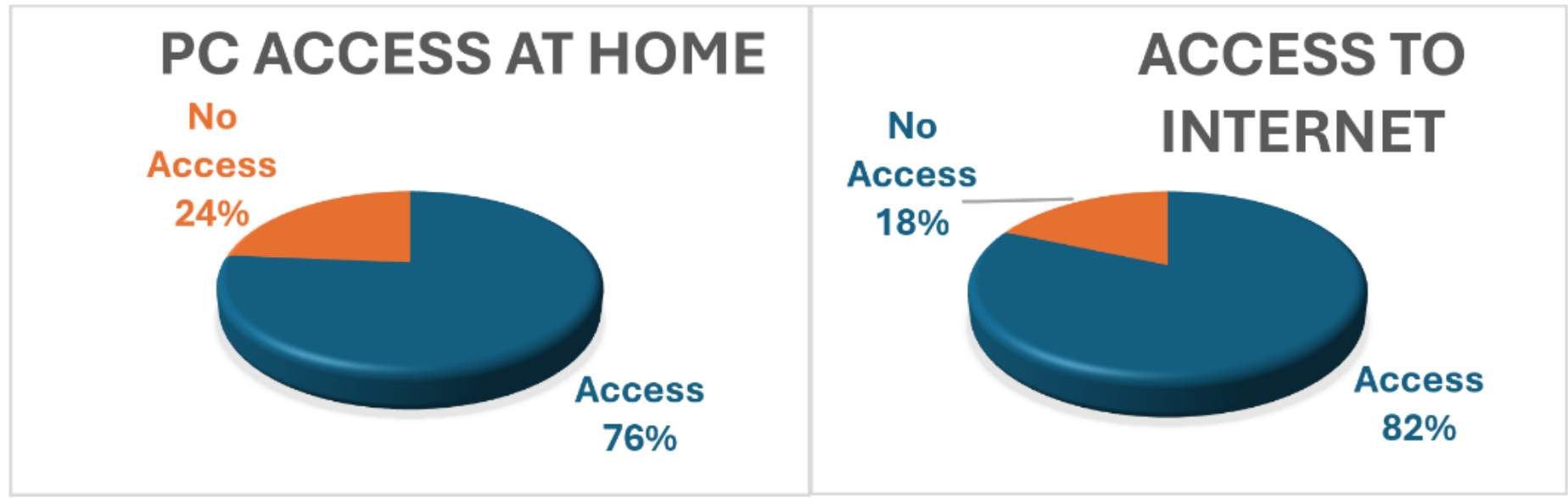

**Fig. 1.** - Access to PC and Internet at home

Figure 1 showed that 29 (about 76%) of the learners have access to a PC at home, while 31 (about 82%) of the learners have access to the internet at home. These findings challenge the narrative that PACs are explicitly mandated to eradicate the digital divide within underserved communities [2]. This reveals that the I-CAN Centre is not predominantly servicing a very underserved area. In addition to these findings, most learners can use the LMS features to access learning content at home, which would help them learn on their own time and at their own pace, a facilitator of self-paced learning.

Fourteen of the learners (about 37%) reported that they do not access the LMS at home. Literature states that LMS provides valuable features that enable learners to access learning content anytime, if they have a mobile or wireless network. This access would enrich their learning experiences and increase learning opportunities [10]. Learner C5L27 stated that the reason for not accessing LMS at home was that the learning activities were easy, and they did not have access to a computer. And learner C2L36 preferred watching YouTube videos that supplement the content on the LMS at home, which does not require access to the LMS. *"The activities are easy because we do them in class because I don't have a computer or laptop at home. But it's enough." C5L27 "Not quite often because I'm using YouTube to learn because the mElimu content is only limited there, you have one topic there, and then with YouTube, you have your variety of topics. Learn more, so I use YouTube more to learn." C2L36*

However, 24 of the learners (about 63%) have stated that they access to LMS at home. However, the frequency of use is different; some mentioned that they access LMS daily, while others said once or more than once a week, and other learners said they access the LMS only if they have work. Seventeen of the learners (about 45%) who access the LMS frequently mentioned that they access the LMS to revise and keep up to date with their work. The finding also shows that learners appreciate accessing the learning content on the LMS during holidays and exams to help them revise their lessons. *"Monday and Wednesday after the course, I will go home and revise, and whenever I have time off, I would like to sit in for 30 minutes to revise my work" C2L3*

Seven learners (about 18%) mentioned that they access the LMS if they have specific learning tasks to complete. For example, they would access the LMS to complete outstanding tasks, prepare for exams, and research a particular topic. *"If we have tasks available, I will go to open to study but not frequently." C5L5 "Only a few times when I study for an exam, and it's only for the exam; exam time, I'll use it not always." C2L29*

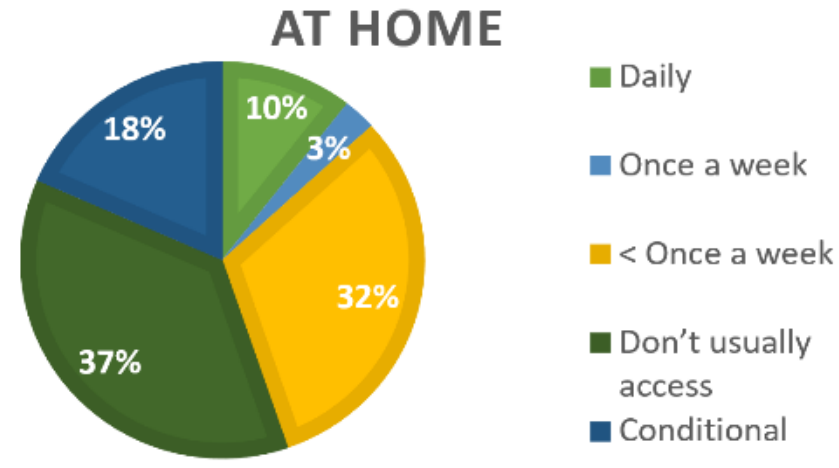

**Fig. 2.** - Frequency of LMS use while at home

### 4.2 Narrative findings and discussion on how LMS facilitates self-paced learning

While all previous descriptive findings responded mainly to the question of familiarity and what the LMS is used for in the context of self-paced learning, the question of how learners use the LMS to facilitate self-paced learning was discussed narratively following an explanatory research approach. It was established through analysis of respondents' transcripts that the LMS has various features designed to facilitate self-paced learning. These also included the management of activities to help learners engage with the work at hand and features that enable learners to take assessments that display grades obtained instantaneously.

When learners were asked how they used LMS to manage their coursework to ensure the completion of their course assessment, findings showed that most learners used the LMS to access the course content. The content contains activities, practice test questions, slides, and videos. Some learners mentioned that they would use the assessment feature, which helped them to retake their tests serval times to help ensure understanding of their work. *"I will try logging on to the website at home and then watch the videos. When I get here, then I will work through everything." C1L22 "Login to geometrics or mElimu, do a few slides or videos, and make a few notes if I have to prepare for a test." C2L3 "Like it just gave me the questions, and like I can make points with continuously so I can always go over it, and it gives me a nice percentage" CL29*

Some learners were not explicit about using the LMS to facilitate learning. They stated that they created a schedule for themselves, set a timetable, created a checklist, and created a priority list for their work. These techniques were outside the system's features, but in conjunction, they facilitated self-paced learning. *"Make a checklist for myself, to know what I've done and then if I learned something; I try to go through my books and watch the YouTube video to ensure I understand the right thing I'm being taught." C5L26 "Basically, I have a schedule, and I stick to the times on the schedule, which helps me keep in line." C2L28*

Following these narratives from the learners, it was established that the LMS presents beneficial opportunities for self-paced learning. While courses could be studied for at home for most learners, the I-CAN centre was integral in that this is the institution where the courses were formerly registered for, and scheduling was done. This formal registration for the courses, together with the availability of the LMS at the I-CAN centre (especially for learners without access at home) responded directly to the question of the opportunities presented to learners by using the LMS to facilitate learning at the I-CAN Centre.

Those having home access to the LMS experienced an added advantage. One of the attractive features of LMS is the accessibility of the application. Some learners stated that they have enjoyed having access to the LMS because it can be accessed from anywhere and using any device. "*mElimu like, not only at the Centre, but I can access it anywhere. So, what's better, like I can do it anywhere" C2L1 "Oh, the fact that you can also use it on your phone, so you don't have to, like, depend on your desktop" C2L3 "I like having access to lessons and books instead of having, like, if I'm sick, maybe I can access it at home and still do my work. Instead of now coming to class and having to do it, I can access it at home. So, it makes it easier also if I can't make it to class" C4L15*

Others expressed that they have found using LMS easy and helpful in facilitating learning; for example, having access to activities, practice test questions, slides, and videos were beneficial to their education. Learners commented on the application's useability; the structure of the courses in the LMS is well-organised and easy to navigate, which makes it a valuable feature to have. *"Everything is there, and it's detailed* *out in the layout, and everything is nice. So, it's a comfortable app to use." C3L34 "That's just the way everything is well constructed. Like you can go through the folders like Photoshop, you can go in, and all the lessons will be there; we get it that way. It's interactive." C4L21 "In just a click of a button, you can find what you need" C4L15 "It's easy; you can easily navigate on the website, and the website isn't broken, or it isn't like delayed." C3L32*

LMS facilitated access to learning resources in a central location. The system also showed progress status on activities, which provided feedback to the learners to evaluate where they were on the learning journey; this enables them to implement this feedback to resubmit the activities to improve their performance. This agrees with the finding by [6] that LMS eases learner management of their course content and access to the course completion, status and certification. Learners mentioned that they prefer accessing the learning content using LMS rather than working through textbooks because of the convenience and the variety of learning content it provides. "*You can see most of what you did wrong and do it over until it is 100%. So, all the answers are there for afterwards." C2L33 "Like it just gave me the questions and like I can make points with my course, so I can always go over it, and it will give me a nice percentage" C2L29 "But now the MElimu. You still get your textbook, but then you get your questions. So, it helps a lot. Like you can do test" C2L38 "it is quicker and easier. Yes. No burning books" C1L20*

Subsequently, learners were asked the things they found most valuable in furthering their studies specifically at the I-CAN Centre. They were asked for motivators of enrolling in courses offered at the I-CAN Centre. Findings showed that learners benefit from having access to learning facilities such as the I-CAN Centre. Some learners mentioned that I-CAN Centre offered bursaries, which allowed them to further their studies. Others said that receiving support from their instructors was valuable and made their learning experience more enjoyable. *"So, I came across that I-CAN Centre; they seem to have the same courses as the other institution. And they provided a bursary."C2L1 "I applied for a bursary, and it was successful. It's 50% off the original fee. It's exciting, and I learn a lot here. Every day it's like another reason why this is the right choice for me." C1L17 "Here (I-CAN Centre), they (Instructors) show you exactly what you should do." C1L 22*

The I-CAN Centre provided an alternative learning opportunity. Some learners had no interest in studying at a traditional learning institution, others could not qualify at a conventional learning institution due to their low grades, and the rest wanted to enrol in a course as they figured out the future. Learners also appreciated being at the I-CAN centre as it offered a conducive and safe learning environment. All these are elements of access that could not be easily reached at other traditional institutions. *"Honestly, I like this more than the traditional version of universities and stuff because, with my ADHD, books don't work for me. …because the Army didn't respond to my application because they have a backlog, so the next best thing for me was to study" C4L12 "I enrolled in other colleges, and they didn't accept me because I scored average in high school. And they didn't accept me, but I wanted to do graphic design. And I wanted to have a further education" C1L16 "But here, they focus on everything. They [instructors] want you to be stress-free. They want you to focus on your work because you're* *doing the work for yourself." C5L27 "Here you get like enough time, and it's even much easier to use because there are not many people around you, so there's no pressure" C5L26*

The study generated the economic freedom and empowerment theme to describe the learning capabilities and ICT skills the learners gained and to get insights into what they would like to achieve upon their course completion. Learners gained various learning capabilities and skills the while studying at the I-CAN Centre. Findings show that the learner's learning capability and ICT skills have significantly improved since enrolling in a course, and these will help them reach their future aspirations. They had improved their computer skills as some did not know how to use a computer before. Learners also gained advanced technical skills and practical knowledge. They learnt how to build a computer on their own, configure applications, troubleshoot the system, and to code programmes. *"To be honest, I knew nothing about computers. Yeah, when I came here, it was all new to me. I didn't know what was in the computer system. since I've been here, I have got a lot of knowledge about how the computer works and what is what of the computer" C2L1 "I guess before, I didn't know some of the applications on Microsoft, and I didn't know certain things. So, I have gained from what I've learned about computers. I used to have like 50% knowledge of computers; now I have like 75% knowledge of computers and how they work and how they do so. I've gained." C3L32 "I can now install Windows on a PC for Windows. Troubleshooting. I can do for Mac and Apple laptops and Windows" C3L31*

Findings also revealed improvements of learners' career aspirations upon completion of their courses. While 34 (about 92%) of the learners would like to find employment upon completing their course, some of the reasons the learners mentioned for finding a job were to find stability, assist in family responsibilities, start their businesses, gain work experience, and to study even further. *"I want to find a stable job, basically, and I've sort of wasted 30 years of my life, and somehow just, I am refocusing my life" C5L13 "To have a good job, they need to be able to provide for my mom and me" C5L8 "…working in the IT field while I'm studying so I can gain experience in the field" C3L31 "I want to design games and start my own company." C1L18 "Further my studies at the university." C5L5*

The social capital and agency theme was derived from understanding the learner's social benefit regarding meeting and connecting with classmates. The study established that the learners had become more self-efficient and confident. They were asked how the social relationships have impacted their learning and lives in general. It was established that the benefits mentioned were intangible. They describe how they felt about established relationships with their peers. Most learners expressed that the most beneficial thing was being able to ask for help from their peers. Collaborations and friendly competition were some of the stated social benefits. This even extended to enrich their cultural diversity and depth. *"When you don't understand something, you can always go to the next person and ask them." C5L5 "We had to go volunteer for two weeks, and we did it together, like, it does help" C5L 7 "Yes, I've made new friends, and I learned from their cultures" C5L10*

Learning at the I-CAN centre directly contributed to their self-efficiency and self-confidence, likewise. Findings reveal that some learners are now self-efficient.

According to [5], self-efficacy is a personal trait describing a learner's ability to successfully perform course activities with facilitated training and technical support to engage effectively. Even though this trait is difficult to measure, some learners said they could manage their work well and independently when asked if they could work without assistance, and self-confidence had improved. *"Yes, I can. It's online. It's computers. It's my type of language" C5L10 "I can work on my own; I also like to help everyone else. I've done this before." C4L21 "I already know what is happening, so I know how to manage my work." C2L1*

### 4.3 Consolidated discussion summary

This study explored Learning Management System (LMS) use at the I-CAN Centre, a Public Access Centre (PAC) in South Africa, focusing on its role in facilitating self-paced learning. While the LMS provided learners with significant benefits, such as improved ICT skills, economic empowerment, and enhanced social capital, these advantages were accompanied by notable challenges. Learners reported that the LMS enabled flexible, self-paced learning supported by accessible infrastructure, affordable internet, and mentorship. The system's features, including progress tracking, multimedia content, and mobile accessibility, contributed to learner engagement and autonomy. Additionally, the I-CAN Centre's community outreach and support mechanisms fostered a sense of agency, confidence, and collaborative learning among participants.

However, several challenges emerged. Learners cited difficulties in understanding course content, navigating complex LMS interfaces, and dealing with outdated materials. Technological barriers such as poor network connectivity, load shedding, and limited digital literacy further hindered effective LMS use. These issues often led to frustration and diminished self-confidence, particularly among learners with low technological readiness. Despite these obstacles, the LMS at I-CAN served as a catalyst for personal and professional development. Learners expressed aspirations for employment, entrepreneurship, and further education, underscoring the transformative potential of PACs in underserved communities. The findings align with existing literature on ICT4D and reinforce the importance of user-centric design, updated content, and robust infrastructural support to maximise the impact of LMS in public access settings.

## 5 CONCLUSION

This study examined the use of a Learning Management System (LMS) at a Public Access Centre (PAC) to facilitate self-paced learning, focusing on learner experiences, opportunities, and challenges. Findings indicate that PACs can deliver affordable, high-quality, and accessible education while fostering ICT skill development and digital inclusion. Even though 95% of learners had no prior LMS experience before starting a programme at I-CAN centre, access to the PAC's LMS enabled flexible learning, with 63% of learners using the system at home.

Despite these benefits, infrastructural limitations, outdated content, and technical challenges persist, indicating the need for continuous instructional re-design and providing learner support when required. While the study aimed to explore self-paced learning, results revealed a strong preference for blended learning, combining LMS use with instructor support. Practically, the research informs PAC management on implementing LMS solutions to expand on education opportunities and other services. However, the study's scope was limited to learner perspectives and a single LMS, suggesting that future research should include other stakeholders and diverse platforms to provide a holistic view of LMS adoption in PACs.

# 6 REFERENCES

[1] T. Priatna, D. S. Maylawati, H. Sugilar, and M. A. Ramdhani, 'Key Success Factors of e-Learning Implementation in Higher Education', *Int. J. Emerg. Technol. Learn.*, vol. 15, no. 17, p. 101, Sep. 2020, doi: 10.3991/ijet.v15i17.14293.

[2] C. Frans and S. Pather, 'Determinants of ICT adoption and uptake at a rural public-access ICT centre: A South African case study', *African Journal of Science, Technology, Innovation and Development*, vol. 14, no. 6, pp. 1575–1590, Sep. 2022, doi: 10.1080/20421338.2021.1975354.

[3] C. Uys and S. Pather, 'A benefits framework for public access ICT4D programmes', *E J Info Sys Dev Countries*, vol. 86, no. 2, p. e12119, Mar. 2020, doi: 10.1002/isd2.12119.

[4] D. M. Thai, D. Duong, M. Falch, C. B. Xuan, and T. T. A. Thu, 'Factors affecting the sustainability of telecentres in developing countries', *Telecommunications Policy*, vol. 46, no. 3, p. 102265, Apr. 2022, doi: 10.1016/j.telpol.2021.102265.

[5] S. F. Dlalisa and D. W. Govender, 'CHALLENGES OF ACCEPTANCE AND USAGE OF A LEARNING MANAGEMENT SYSTEM AMONGST ACADEMICS', *International Journal of eBusiness and eGovernment Studies*, pp. 63–78, Jan. 2020, doi: 10.34111/ijebeg.202012105.

[6] P. Kumar and S. Sridhar, 'Review Study on E-Learning in Higher Education Administration and Management', *International Journal of Innovative Technology and Research*, vol. 8, no. 2, pp. 9506–9511, 2020.

[7] M. A. Almaiah, A. Al-Khasawneh, and A. Althunibat, 'Exploring the critical challenges and factors influencing the E-learning system usage during COVID-19 pandemic', *Educ Inf Technol*, vol. 25, no. 6, pp. 5261–5280, Nov. 2020, doi: 10.1007/s10639-020-10219-y.

[8] E. Kagona and I. B. Lewis, 'AI-DRIVEN EXAMS CLEARANCE SCHEME AS A PART OF THE EXISTING E-LEARNING SYSTEMS: CASE STUDY (THE INTERNATIONAL UNIVERSITY OF EAST AFRICA, AND MAKERERE UNIVERSITY)', 2022.

[9] A. Aldiab, H. Chowdhury, A. Kootsookos, F. Alam, and H. Allhibi, 'Utilization of Learning Management Systems (LMSs) in higher education system: A case review for Saudi Arabia', *Energy Procedia*, vol. 160, pp. 731–737, Feb. 2019, doi: 10.1016/j.egypro.2019.02.186.

[10] Z. Gashi Shatri, 'Advantages and Disadvantages of Using Information Technology in Learning Process of Students', *tused*, p. 3, Sep. 2020, doi: 10.36681/tused.2020.36.

[11] R. F. Kassongo, W. D. Tucker, and S. Pather, 'Government facilitated access to ICTs: Adoption, use and impact on the well-being of indigent South Africans', in *2018 IST-Africa Week Conference (IST-Africa)*, IEEE, 2018, p. Page 1 of 10-Page 10 of 10.

[12] S. Pokhrel and R. Chhetri, 'A Literature Review on Impact of COVID-19 Pandemic on Teaching and Learning', *Higher Education for the Future*, vol. 8, no. 1, pp. 133–141, Jan. 2021, doi: 10.1177/2347631120983481.

[13] S. L. Booi, W. Chigona, P. Maliwichi, and K. Kunene, 'The Influence of Telecentres on the Economic Empowerment of the Youth in Disadvantaged Communities of South Africa', in *Information and Communication Technologies for Development. Strengthening Southern-Driven Cooperation as a Catalyst for ICT4D*, vol. 551, P. Nielsen and H. C. Kimaro, Eds, in IFIP Advances in Information and Communication Technology, vol. 551. , Cham: Springer International Publishing, 2019, pp. 152–167. doi: 10.1007/978-3-030-18400-1_13.

[14] S. Geng, K. M. Y. Law, and B. Niu, 'Investigating self-directed learning and technology readiness in blending learning environment', *Int J Educ Technol High Educ*, vol. 16, no. 1, p. 17, Dec. 2019, doi: 10.1186/s41239-019-0147-0.

[15] R. Castro, 'Blended learning in higher education: Trends and capabilities', *Educ Inf Technol*, vol. 24, no. 4, pp. 2523–2546, Jul. 2019, doi: 10.1007/s10639-019-09886-3.

[16] E. Ebbini, 'Enhancing collaborative and self-paced learning in traditional and distance education settings', in *Higher education in the Arab world: E-learning and distance education*, Springer, 2023, pp. 223–245.

[17] M. N. K. Saunders, P. Lewis, and A. Thornhill, *Research methods for business studends*, Ninth Edition. New York, NY: Pearson, 2023.

[18] A. Coates, 'The prevalence of philosophical assumptions described in mixed methods research in education', *Journal of Mixed Methods Research*, vol. 15, no. 2, pp. 171–189, 2021.

[19] A. Musundire, 'Understanding the Research Onion and Its Application in Educational Leadership and Management Research: Making Use of Saunders' Research Model', in *Research Methods for Educational Leadership and Management*, IGI Global Scientific Publishing, 2025, pp. 355–384.

[20] J. Sudrajat, A. Mayasari, and O. Arifudin, 'Enhancing the quality of learning through an e-learning-based academic management information system at Madrasah Aliyah Negeri', *EDUKASIA Jurnal Pendidikan dan Pembelajaran*, vol. 5, no. 2, pp. 621–632, 2024.

[21] E. Knott, A. H. Rao, K. Summers, and C. Teeger, 'Interviews in the social sciences', *Nat Rev Methods Primers*, vol. 2, no. 1, p. 73, Sep. 2022, doi: 10.1038/s43586-022-00150-6.

[22] V. Braun and V. Clarke, 'Thematic analysis.', in *APA handbook of research methods in psychology, Vol 2: Research designs: Quantitative, qualitative, neuropsychological, and biological.*, H. Cooper, P. M. Camic, D. L. Long, A. T. Panter, D. Rindskopf, and K. J. Sher, Eds, Washington: American Psychological Association, 2012, pp. 57–71. doi: 10.1037/13620-004.